\documentclass[conference,10pt]{IEEEtran}
\usepackage[top=54pt, left=54pt, right=54pt, bottom= 0.95in]{geometry}

\usepackage{tabularx}
\usepackage{cite}
\usepackage{etoolbox}
\makeatletter
\patchcmd{\@makecaption}
  {\scshape}
  {}
  {}
  {}
\makeatother

\usepackage{algorithm}
\usepackage{algpseudocode}
\usepackage{cite}
\ifCLASSINFOpdf
   \usepackage[pdftex]{graphicx}
\else
   \usepackage[dvips]{graphicx}
\fi

\usepackage{amsmath}
\usepackage{array}
\usepackage{float}
\usepackage{url}

\begin{document}

\title{RL-based Relay Selection for Cooperative WSNs in the Presence of Bursty Impulsive Noise}

\author{
    \IEEEauthorblockN{Hazem Barka\IEEEauthorrefmark{1},
    Md Sahabul Alam\IEEEauthorrefmark{2},
    Georges Kaddoum\IEEEauthorrefmark{1},
    Minh Au\IEEEauthorrefmark{3},
    and Basile L. Agba\IEEEauthorrefmark{3}}
    \IEEEauthorblockA{\IEEEauthorrefmark{1}École de Technologie Supérieure (ETS), Montreal, Canada
    \\hazem.barka.1@ens.etsmtl.ca, georges.kaddoum@etsmtl.ca}
    \IEEEauthorblockA{\IEEEauthorrefmark{2}California State University, Northridge, California, USA
    \\md-sahabul.alam@csun.edu}
    \IEEEauthorblockA{\IEEEauthorrefmark{3}Hydro-Quebec Research Institute (IREQ), Varennes, Canada
    \\au.minh2@hydroquebec.com, agba.basilel@hydroquebec.com}}

\maketitle

\begin{abstract} 
The problem of relay selection is pivotal in the realm of cooperative communication. However, this issue has not been thoroughly examined, particularly when the background noise is assumed to possess an impulsive characteristic with consistent memory as observed in smart grid communications and some other wireless communication scenarios. In this paper, we investigate the impact of this specific type of noise on the performance of cooperative Wireless Sensor Networks (WSNs) with the Decode and Forward (DF) relaying scheme, considering Symbol-Error-Rate (SER) and battery power consumption fairness across all nodes as the performance metrics. 
We introduce two innovative relay selection methods that depend on noise state detection and the residual battery power of each relay. The first method encompasses the adaptation of the \textit{Max-Min} criterion to this specific context, whereas the second employs Reinforcement Learning (RL) to surmount this challenge. Our empirical outcomes demonstrate that the impacts of bursty impulsive noise on the SER performance can be effectively mitigated and that a balance in battery power consumption among all nodes can be established using the proposed methods.


\end{abstract}

\section{Introduction}
Cooperative communication has emerged as a powerful tool in the wireless communication realm, offering significant benefits in terms of system performance, spatial diversity, and reliability \cite{novel_BRS_alam}. Primarily, it allows single-antenna wireless devices to collaboratively share their antennas and resources, mimicking the capabilities of a Multiple-Input Multiple-Output (MIMO) system without the physical constraints and cost implications. One of the foundational components of cooperative communication is relay selection, wherein the optimal relay node is chosen to retransmit a signal, optimizing an objective function under specified constraints. Relay selection has been explored extensively for traditional communication systems operating under Additive White Gaussian Noise (AWGN) environments \cite{energy_consumption, energy_consumption_faireness,awgn_RS_2,awgn_RS_3,time_based_relay_selection_alam,time_based_relay_selection_original}. 

However, real-world applications, such as smart grid communications, introduce unique challenges. Smart grids, a result of rising electricity demands, represent the next generation of power networks by integrating bidirectional power flow and self-healing capabilities with communication and information technologies \cite{measurements, smart_grids}. In this environment, Wireless Sensor Networks (WSNs) play a crucial role due to their inherent flexibility and cost-effectiveness. Yet, WSNs in smart grids often encounter non-Gaussian, impulsive noise, especially around power transmission lines and substations. This noise severely affects the quality of service metrics, notably the Symbol-Error-Rate (SER) or Bit-Error-Rate (BER).

Aside from smart grid applications, impulsive noise environments are also observed in areas such as vehicular communication systems, underwater acoustic communication networks, and industrial control systems \cite{IN_is_everwhere}. Given the prevalence of such noise characteristics, it becomes imperative to design relay selection schemes for cooperative WSNs tailored to handle such noise scenarios.

 To our knowledge, there are no major studies that focused on scenarios involving impulsive noise with memory, which is common in smart grids \cite{measurements,mitigation_techniques,novel_BRS_alam}, except \cite{novel_BRS_alam}. Moreover, implementing the method proposed in \cite{novel_BRS_alam} in real-life scenarios is challenging due to its reliance on a large number of assumptions and its disregard for battery life constraints within the optimization problem.

Given these considerations, formulating an appropriate relay selection method based on traditional optimization methods could be challenging. Thus, incorporating a certain pattern recognition-based behavior into the selection criterion would be an appropriate approach. In this regard, related studies such as \cite{DQN}, \cite{MAB_energy}, and \cite{Actor_critic} have utilized Reinforcement Learning (RL) to optimize a future reward (which could include mutual information, energy consumption, etc.) by leveraging environment patterns. However, despite their promising results, none of these studies have considered impulsive noise.
In this paper, we delve into the effects of bursty impulsive noise on WSNs utilizing the Decode and Forward (DF) scheme, with a specific emphasis on SER and power consumption fairness. Our objective is to forge innovative relay selection strategies tailored to these challenging noise conditions.

Toward this end, we introduce two distinct relay selection techniques. The first adapts the \textit{Max-Min} criterion from \cite{novel_BRS_alam} for our noise scenario, incorporating noise state detection and residual battery power of relays. The second harnesses RL techniques, a pioneering approach in this context, to counter the challenges posed by impulsive noise. Preliminary empirical results affirm the efficacy of both methods in achieving robust performance and power consumption equity in the face of bursty impulsive noise.

The rest of the paper is organized as follows: In Section II, we introduce the system model. Section III investigates the impact of bursty impulsive noise on the \textit{Max-Min} method. Section IV is dedicated to exploring the effectiveness of RL for addressing the relay selection problem in the presence of bursty impulsive noise. The last section serves as the paper's conclusion.

\section{System Model}
\subsection{Signal Model}
This section considers a DF cooperative network with $M$ relays assisting in data transfer between a Source-Destination (SD) pair. We assume that all node terminals share a single communication channel, possess a single transmit/receive antenna, and operate in a half-duplex mode.

The transmission process is divided into two distinct time slots. In the first slot, the source transmits data to the destination and other nodes. Subsequently, in the second time slot, only the selected relay decodes the message received from the source and forwards it to the destination. During this period, the source remains silent. Following this, the destination combines the received noisy sequences using Maximum Ratio Combining (MRC).

We consider a frame-by-frame relay selection. For each frame with size $K$, the source $S$ generates a Quadrature Phase Shift Keying (QPSK) sequence, $\left(x_{S, 0}, \ldots, x_{S, K-1}\right)$. The modulated signal is then broadcasted to the destination $D$ and the relay nodes $\left\{R_{1}, R_{2}, \ldots, R_{M}\right\}$. The signals received at the relay $R_{m}$ and $D$ at each time epoch $k, k=0,1, \ldots, K-1$ can be written, respectively,
as
\begin{equation}
    y_{S R_{m}, k} =\sqrt{P_{S}} h_{S R_{m}, k} x_{S, k}+n_{S R_{m}, k}, 
\end{equation}
\begin{equation}
y_{S D, k} =\sqrt{P_{S}} h_{S D, k} x_{S, k}+n_{S D, k},
\end{equation}

where $P_{S}$ is the average source transmission power per symbol, $x_{S, k}$ is the transmitted symbol from $S$, $h_{i j, k}$ is the $i j$ link channel coefficient, $i$ $\in$ $(S, R_m)$ and j $\in$ $(R_m, D)$, and $n_{i j, k}$ is the noise sample.\\
Using MRC, the received signal at the destination can be obtained as:
\begin{equation}
y_{D, k} = \frac{ H_k^\dag \times Y_k}{||H_k||},
\end{equation}
where $H_k = [h_{S R_{m}, k},h_{S D, k}]$ and $Y_k = [y_{S D, k},y_{R_{m} D, k}]$.

In this work, we presume an ideal DF cooperation protocol where the relay is capable of discerning whether the transmitted symbol has been decoded correctly or not.
\subsection{Assumptions on the Channel Coefficients}
Regarding the fading, we assume that the channel coefficients of each $i j$ link follow a Rayleigh distribution. The coherence time is equal to one frame duration. As such, $h_{i j}$ is modeled as a zero-mean, independent, circularly symmetric complex Gaussian random variable with variance  $\sigma_{i j}^2 \equiv \mathrm{E}\left\{\left|h_{i j}\right|^{2}\right\}=1 / \lambda_{i j}^{\eta}$, where $\lambda_{i j} = d_{ij}/d_{SD}$ denotes the relative distance from $i$ to $j$, $d_{ij}$ signifies the actual distance from $i$ to $j$, and $\eta$ represents the path loss exponent. The destination node assumes perfect Channel State Information (CSI) related to all links in the WSN cluster, which is feasible since we are considering a slow-fading scenario.

\subsection{Assumptions on the Noise}
Upon investigation, it is determined that the links between the relays and the destination should not be affected by impulsive noise, given that the destination is located outside the measurement field and is typically connected to the monitoring center \cite{novel_BRS_alam}. Consequently, $n_{R_{m} D, k}$ follows a complex Gaussian distribution. However, it is assumed that the noise samples $n_{S R_{m} , k}$ are impulsive. Furthermore, both the channel coefficients and the noise samples for each link are statistically independent.

Over the past decades, numerous models have been utilized to characterize impulsive noise \cite{IN_models}. In our work, we employ the Two-state Markov Gaussian model (TSMG) to characterize bursty impulsive noise. The model consists of a Markov space composed of two states: a good state, $G$, when the transmitted signal is affected only by the background AWGN, and a bad state, $B$, when we have impulsive interferers.  For more details about the TSMG model, please check \cite{novel_BRS_alam} and \cite{mitigation_techniques}. 



\section{The Max-Min relay Selection Criterion}
\subsection{Impact of Bursty Impulsive Noise on the Conventional Max-Min Criterion}

\begin{figure}[t]
    \centering
   \includegraphics[width=0.65\linewidth]{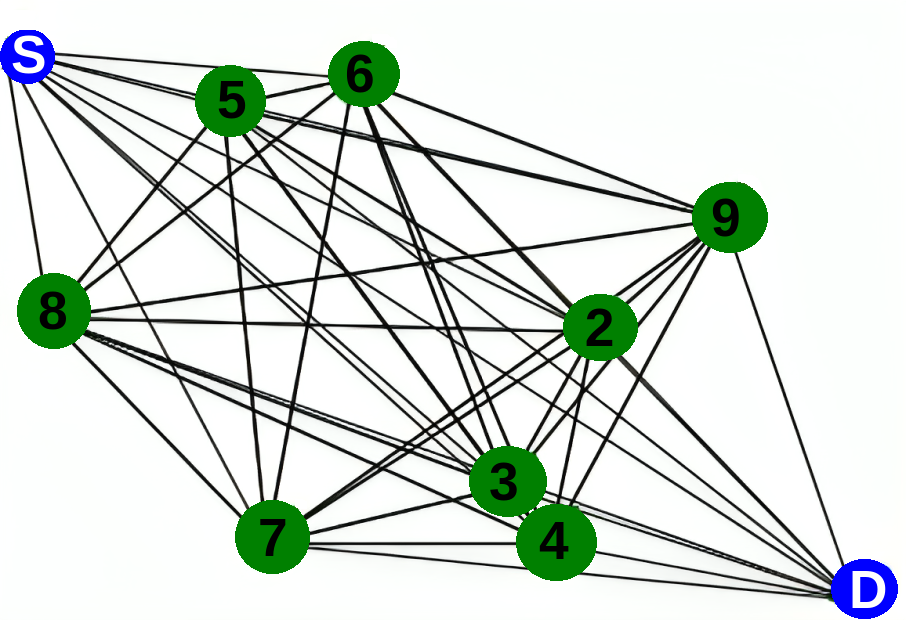}
    \caption{Illustration of the sensors' repartition.}
        \label{fig:simulated_env}
\end{figure}

The popular method for best relay selection in AWGN environments is the conventional max-min criterion \cite{min-max} : 
\begin{equation}
R_{s}=\underset{m \in\{1,2, \ldots, M\}}{\arg \max }\left\{\min \left\{\left|h_{S R_{m}}\right|^{2},\left|h_{R_{m} D}\right|^{2}\right\}\right\},
\end{equation}
where $R_s$ is the selected relay. This selection criterion is considered the best method in terms of SER for AWGN environments. Therefore, we will investigate its performance under bursty impulsive noise. The parameters used for the simulation are detailed in Table \ref{tab:params}. The sensor nodes are positioned according to a two-dimensional Poisson distribution, with the exception of the source and destination nodes ($S = 1$ and $D$), which are positioned at the corners of the field, as illustrated by Fig. \ref{fig:simulated_env}.

\begin{table}[t]
\caption{\label{tab:params}The simulation settings.}
\begin{tabular}{ |p{6cm}|p{1.7cm}|}
 \hline
 Parameter & Value\\
 \hline Number of nodes & 10    \\
  \hline Frame length $K$ / Coherence Time $T_c$ & 1000 symbols  \\
  \hline Number of symbols per one curve point  & $10000 \times 10$   \\
  \hline Impulsive noise memory $\gamma$ & 100  \\
  \hline Bad to good states noise power ratio $R = \sigma_B^2/ \sigma_G^2$ & 100  \\
  \hline The stationary probability $P_B$ & 0.1  \\
    \hline The path loss exponent  $\eta$ & 2  \\
 \hline
\end{tabular}
\end{table}

\begin{figure}[t]
    \centering
   \includegraphics[width=0.8\linewidth]{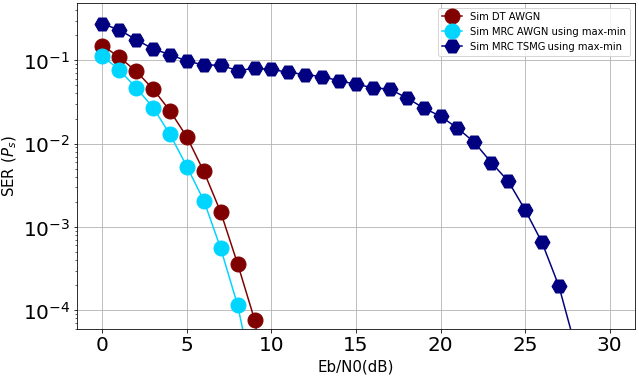}
    \caption{SER performance using the conventional max-min criterion for a QPSK modulated symbols under TSMG and AWGN noise and slow Rayleigh fading.}
    \label{IN_effect_SER}

\end{figure}
\begin{figure}[t]
    \centering
   \includegraphics[width=0.8\linewidth]{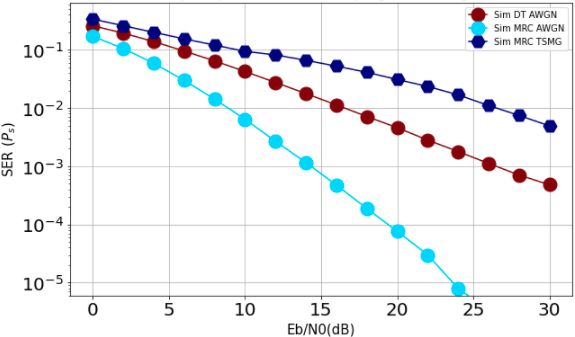}
    \caption{SER performance using the conventional \textit{Max-Min} criterion for a QPSK modulated symbol under TSMG and AWGN noise and fast Rayleigh fading ($T_c = T_s$).}
    \label{IN_effect_SER_FF}

\end{figure}

 As shown in Fig. \ref{IN_effect_SER}, when the noise is Gaussian, cooperative relaying using the MRC results in better performance than Direct Transmission (DT). However, the slope observation reveals that despite the presence of two independently faded versions of the signal, the diversity order does not increase significantly. This can be attributed to the fact that $y_{R_m D,k}$ does not include all symbols because the relay has discarded some of them. These symbols are expected to be falsely detected with high probability; hence they are removed. Furthermore, the impact of slow fading is also a contributing factor, as the diversity gain is more noticeable in Fig. \ref{IN_effect_SER_FF}.

\begin{figure}[t]
    \centering
   \includegraphics[width=0.8\linewidth]{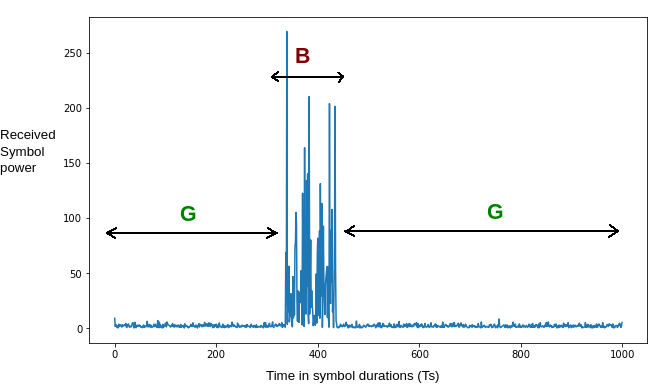}
    \caption{The received signal power by symbol $y_{S R_{m}, k}^2$ for a frame of size 1000 symbols under the considered bursty impulsive noise scenario.}
    \label{signal_power}
\end{figure}

Overall, MRC consistently yields improved SER. It is worth noting that by increasing the number of selected relays, the diversity gain becomes more pronounced. Regarding the case of TSMG noise, it is clear that the TSMG noise leads to significant performance degradation compared to the AWGN case \cite{novel_BRS_alam}. This can be attributed to the high noise power in the bad state, as illustrated in Fig. \ref{signal_power}, where the effect of impulsive interferers is quite prominent.
In this situation, the ability of the relay to avoid transmitting symbols that are possibly incorrectly detected is compromised. This is due to the uncertainty surrounding whether the noise power is $\sigma_G^2$ or $\sigma_B^2$. Consequently, the detection of the noise state becomes a necessity, and it should be incorporated into the relay selection process.

\begin{figure}[t]
    \centering
   \includegraphics[width=0.78\linewidth]{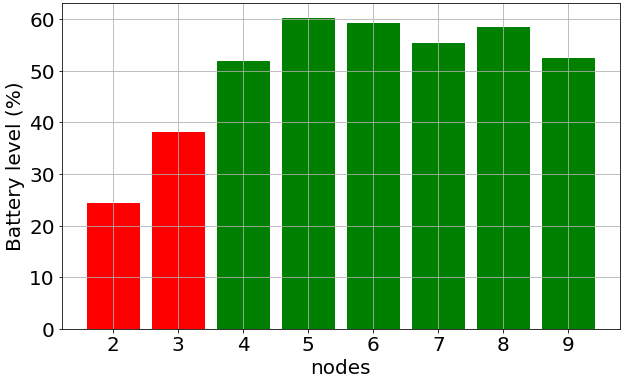}
    \caption{Remaining battery level for each relay node after the simulated communication.}
    \label{IN_effect_Battery}
\end{figure}

In addition, Fig. \ref{IN_effect_Battery} illustrates that certain nodes (specifically $2$ and $3$) may deplete their batteries significantly earlier than others. These nodes are situated centrally along the path between the source and destination, making them optimal candidates for selection under the \textit{Max-Min} criterion, given that the channel coefficient is distance-dependent.

\begin{algorithm}[t]
\caption{The Proposed \textit{Max-Min} relay selection criterion}\label{alg:cap}
\begin{algorithmic}
\Require For each frame at each relay $R_m$ :
\begin{itemize}
    \item  A noise state detector to estimate the percentage of having bad states $P_{B,R_m}$.
    \item Possible error estimator to evaluate if the received symbol can be wrongly detected or not in the case of having a good state.
\end{itemize}

 \For{each frame sent from $S$} 
 \begin{enumerate}
    \item Evaluate $P_{B,R_m}$ for each relay $R_m$ and send it to the destination node.
    
    \item Select a subset $\Omega$ composed from the best three relays in terms of having the lowest $P_{B,R_m}$.
    \item Select the suitable relay $R_{s}$:
        \end{enumerate} 
        
        \begin{equation}
 R_{s}  \gets \underset{R_m \in \Omega}{\arg \max }\left\{\min \left\{\left|h_{S R_{m}}\right|^{2},\left|h_{R_{m} D}\right|^{2}\right\}\cdot \alpha_m \right\},
\end{equation}
    where $\alpha_m $ is a penalizing coefficient related to the battery power consumption:
   \begin{equation}
 \alpha_m \gets \frac{g(R_m) - \underset{n\in\{1,\ldots,M\}}{\min g(R_n)}}{\underset{n\in\{1,\ldots,M\}}{\max g(R_n)} - \underset{n\in\{1,\ldots,M\}}{\min g(R_n)}},
   \end{equation}
and $g$ is a function that characterizes how much battery power is available for every relay.

4) $R_{s}$ makes the transmission without sending the symbols that are in state $B$ or that have a high probability of being wrongly detected if the state is $G$.

5) Combine the received signals using MRC.
\EndFor
\end{algorithmic}
\end{algorithm}

\subsection{Proposed Max-Min Relay Selection Criterion Adapted to Bursty Impulsive Noise Environments with Battery Fairness}
In this section, we elaborate on a modified version of the \textit{Max-Min} relay selection criterion that takes into account noise state information (the percentage of $G$ or $B$ states) as well as the remaining battery life of each relay.

\subsubsection{Method}

Incorporating a noise state detector ($G$ or $B$) is vital to prevent the transmission of symbols corrupted by impulsive interference. Fortunately, numerous methodologies have been proposed in the literature to tackle this challenge, notably \cite{novel_BRS_alam} and \cite{lstm_based_state_detection}. With the help of a noise state detector, we introduce a new method, as illustrated by Algorithm \ref{alg:cap}.
\begin{figure}[t]
    \centering
   \includegraphics[width=0.75\linewidth]{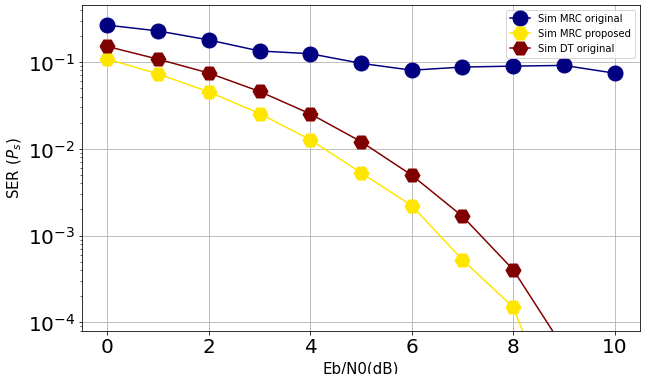}
    \caption{SER performance using the proposed \textit{Max-Min} criterion for a QPSK modulated symbol under TSMG and AWGN noise and slow Rayleigh fading.}
    \label{SER_proposed}
\end{figure}

\begin{figure}[t]
    \centering
   \includegraphics[width=0.8\linewidth]{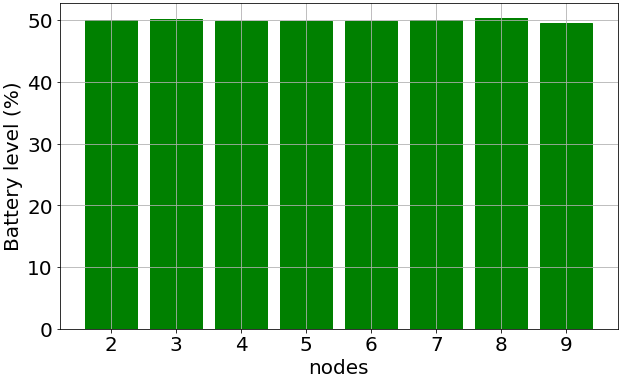}
    \caption{Remaining battery level for each relay node after the simulated communication using the proposed method.}
    \label{Battery_proposed}
\end{figure}

\subsubsection{Numerical Results and Interpretations}
We test the proposed method using the same settings as in the previous section. For noise state detection, we assume the genie condition, which means we have perfect knowledge about the noise states. Fig. \ref{SER_proposed} demonstrates that the proposed method outperforms the original \textit{Max-Min} criterion, which does not take into account the bursty and impulsive nature of the noise. By avoiding the relays affected by impulsive noise, we can achieve near AWGN SER performance. In addition to the improvement in SER, we also succeeded in ensuring fairness in battery consumption between all the relays. As shown in Fig. \ref{Battery_proposed}, there is almost equal remaining battery power among all the relays (2 to 9). This is due to the penalization coefficient $\alpha_m$, which increases the probability of selection for nodes with higher remaining battery power.

\section{Reinforcement Learning Based Relay Selection in the Presence of Bursty Impulsive Noise}

 This section presents an RL approach to solve the relay selection problem in WSNs under bursty impulsive noise conditions. We formulate the problem as a Markov Decision Process (MDP) and utilize RL techniques to identify the optimal policy. RL has gained popularity for its ability to optimize non-convex problems, which are challenging to address via traditional methods. In RL, an agent is created to execute specific actions within a predefined environment, often characterized by single or multiple states, modeled using MDP \cite{RL_book,RL_physical_layer}. The agent's goal is to maximize the cumulative reward over time. Applying this to our case, we have:
 \begin{figure}[t]
    \centering
   \includegraphics[width=0.8\linewidth]{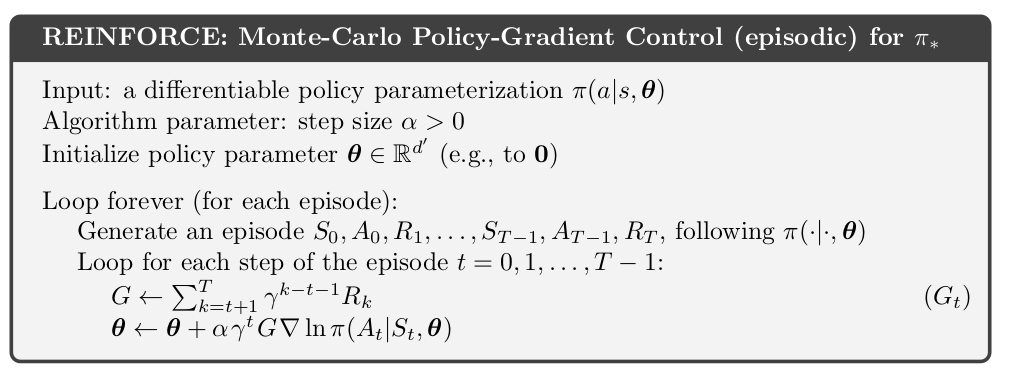}
    \caption{The REINFORCE algorithm \cite{RL_book}.}
    \label{reinforce}
\end{figure}

\begin{itemize}
\item \textbf{States:} For every frame, the state representation of our environment is a multi-feature vector, encompassing elements drawn from the channel gains of all WSN links, the proportion of impulses relative to the overall frame length $P_B$ at every relay, and the residual battery power for each relay. 

\item \textbf{Reward:} This attribute should be based on the metric we aim to optimize, such as the SER difference. It can be calculated as the deviation between the actual SER and the optimal SER determined using the \textit{Max-Min} criterion under AWGN conditions. However, it is advised not to include the remaining battery energy as part of the reward to avoid potential instability issues caused by optimizing two different metrics simultaneously.

\item \textbf{Actions:} The actions in our RL model are straightforward - each action corresponds to the selection of a relay.

\end{itemize}

It should be noted that state transitions rely solely on the remaining battery power as other state features are random.

Choosing a suitable RL algorithm requires considering our model's specifics. Our model dynamics, excluding sensor battery life, are mostly random and unlearnable, necessitating a model-free method. Tabular methods and Value Function Approximation (VFA) are not feasible due to our infinite state space and our focus on estimating rewards for specific actions at given states rather than solely on states. Thus, we have three viable options:
\begin{itemize}

\item Deep Q-Learning (DQN): This method estimates the expected cumulative reward of a pair (state, action).

\item Policy Gradient (PG): This method involves directly estimating the suitable action given a specific state.

\item Actor-Critic (AC): This method directly estimates the suitable action given a specific state in addition to its value function to enhance the stability compared to PG.
\end{itemize}

Although all three categories of algorithms are suitable for our case, we chose a PG method. This decision was made because AC methods tend to be more complex, and DQN introduces an unnecessary step. Among PG algorithms, REINFORCE stands out as the most prominent and is detailed in Fig. \ref{reinforce}. Our objective is to train a Deep Neural Network (DNN) characterized by a weight matrix \( \boldsymbol{\theta} \), known as the policy network. This network assesses the probability of choosing each action \( a_k \) given a state \( s_k \), denoted by \( \pi(a_k|s_k, \boldsymbol{\theta}) \).

\subsection{The Proposed Method}

Our goal is to minimize the SER while ensuring equitable battery consumption among the nodes. These two distinct tasks need to be separated during the learning process to optimize the training stability, a common issue with Monte Carlo-based RL methods like REINFORCE. The approach we have chosen involves creating separate agents for each task. Alternatively, we could have utilized a learning-based approach for one of the tasks. However, in this case, we have opted for a threshold-based validation method to determine the eligibility of a relay for transmission and whether to proceed to the next relay in line. Specifically, an eligible relay must have a remaining battery power above a certain threshold relative to the minimum remaining battery power among all the relays. The details of this method are shown in Algorithm \ref{RL_proposed}.

\begin{algorithm}[H]
\caption{The proposed RL-based relay selection method}\label{RL_proposed}
\begin{algorithmic}
\Require 

 A noise state detector to estimate the percentage of having bad states $P_{B, R_m}$, and a possible error estimator to evaluate if the received symbol can be wrongly detected.

 \For{each frame sent at $k$ from the source node $S$} 
 \begin{enumerate}
    \item Evaluate $P_{B,R_m}$ for each $R_m$ and send it to $D$.
    \item At the destination node $D$, the policy network evaluates $\pi(a_k = R_m|s_k, \boldsymbol{\theta})$ for each relay $R_m$.
    \item Ranking all the relays $R_m$ in a list $M$ from the maximum to the minimum in terms of $\pi(a_k = R_m|s_k, \boldsymbol{\theta})$.
 \item \end{enumerate}
     \For{each $R_m$ in $M$} \\
            \hspace{ 20pt}i) Calculate: 
                 \begin{equation}
                \alpha_m = \frac{ g(R_m) - \underset{n\in\{1,\ldots,M\}}{\min g(R_n)}  }{\underset{n\in\{1,\ldots,M\}}{\max g(R_n)}},
            \end{equation}
        \\\hspace{ 20pt} where $g$ is a function that characterizes the remaining  \\\hspace{ 20pt}battery power for every relay.\\
              \hspace{ 20pt}ii) Check:
            \begin{equation}
    \alpha_m > \beta \cdot \frac{ \underset{n\in\{1,\ldots,M\}}{\max g(R_n)} - \underset{n\in\{1,\ldots,M\}}{\min g(R_n)}  }{\underset{n\in\{1,\ldots,M\}}{\max g(R_n)}},
                \end{equation}
     where $\beta$ is a scaling coefficient. If the condition is valid \\\hspace{ 20pt} then the selected relay $R_{s}$ is $R_m$ and stop the loop.  \\\hspace{ 20pt}Otherwise, move to the next $R_m$.
\EndFor

 \begin{enumerate}
   \setcounter{enumi}{4}
    \item $R_{s}$ transmits without sending the symbols that are in state $B$ or that have a high probability of being wrongly detected if the state is $G$.

 \item $D$ Combines the received signals using MRC.
 \item Gathering the reward:           
 \begin{equation}
R_{k+1} = - \lambda \cdot (SER_k^{obt} - SER_k^{opt} )+ \mu, 
    \end{equation}

 where $SER_k^{obt}$ is the obtained SER, $SER_k^{opt}$ is the optimal SER (Calculated using the \textit{Max-Min} criterion under AWGN instead of TSMG noise), $\mu$ is a reward offset, and $\lambda$ is a scaling coefficient.
 \item Store the experience in the agent memory (also called Buffer Replay).
\end{enumerate}
\EndFor

The policy network with parameters $\boldsymbol{\theta}$ should be trained whenever a dataset with size $T$ is gathered using:
 \begin{equation}
\boldsymbol{\theta} \leftarrow \boldsymbol{\theta}-\alpha  \sum_{t = 0}^{T-1} R_{t+1} \nabla \ln \pi\left(a_{t} \mid s_{t}, \boldsymbol{\theta}\right).
    \end{equation}

\end{algorithmic}
\end{algorithm}

\begin{figure}[t]
    \centering
   \includegraphics[width=0.75\linewidth]{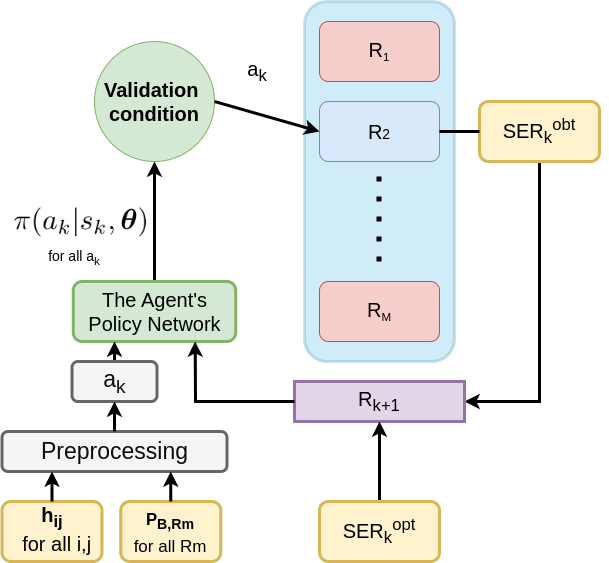}
    \caption{Illustration of the proposed RL relay selection method.}
    \label{RL_relay_selection}
\end{figure}

\begin{figure}[t]
    \centering
   \includegraphics[width=0.9\linewidth]{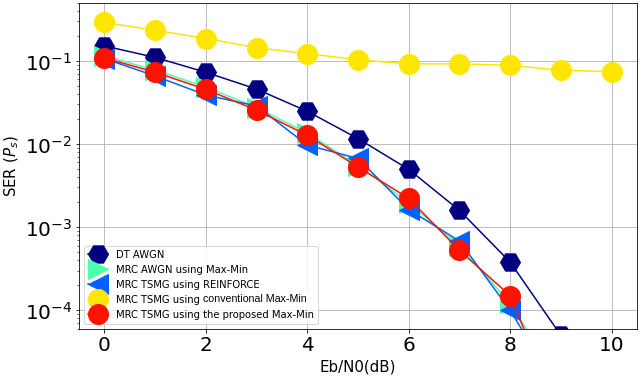}
    \caption{SER comparison for a QPSK modulated symbol under TSMG and AWGN noise and slow Rayleigh fading.}
    \label{SER_comparison}
\end{figure}
\subsection{Numerical Results and Interpretations}
In order to generate the results presented in Fig. \ref{SER_comparison}, we trained the RL agent for various $E_b/N_o$ values ($E_b$ is the energy per bit and $N_o$ is the Good state noise power spectral density). The training phase exhibited some initial instabilities with high variance, which is common for RL in complex environments; however, the model eventually converged to its optimal performance.

 As depicted in Fig. \ref{SER_comparison}, our proposed RL method achieves the same SER performance as the \textit{Max-Min} criterion under AWGN, as well as our proposed modified \textit{Max-Min} criterion under the TSMG noise model. Regarding battery fairness, the RL-based method demonstrates a similar behavior to that illustrated in Fig. \ref{Battery_proposed}. It is important to note that our method was tested using randomly generated channel gains and noise samples. However, the advantage of our RL approach becomes apparent when patterns exist within the environment, particularly in the case of impulsive noise samples.

\section{Conclusion}
This work emphasized the significance of relay selection in WSNs amidst bursty impulsive noise. We examined its impact on SER performance and battery consumption and modified the popular \textit{Max-Min} criterion to balance SER optimization and battery consumption fairness. We also introduced a novel RL-based approach for the relay selection problem in such noise environments. Experiments showed that both our modified \textit{Max-Min} and RL-based methods achieved the desired performance and battery fairness.

Importantly, these methods can be generalized to multiple relay selection scenarios and can incorporate WSN-specific constraints, enabling more comprehensive solutions.

\bibliography{main.bbl}
\IEEEpeerreviewmaketitle

\end{document}